\documentclass[10pt,letterpaper]{article}
\usepackage{amsmath}
\usepackage{amssymb}
\usepackage{cite}
\usepackage{color}
\usepackage{opex3}
\usepackage{txfonts}
\usepackage{epstopdf}
\usepackage{amssymb}
\usepackage{tipa}
\usepackage{graphicx}
\usepackage{txfonts}
\usepackage{amssymb}
\usepackage{extarrows}
\usepackage{float}
\usepackage{amsfonts}
\usepackage{inputenc}

\begin{document}
\title{Tunable photon blockade in a two-mode second-order nonlinear system embedded with a two-level atom\\}
\author{Hongyu Lin$^{1,2}$, Xiangyi Luo$^{2}$, Yanhui Zhou$^{3}$, Hui Yang$^{1}$ and Zhihai Yao$^{1*}$ \\}
\address{$^1$ Department of Physics, Changchun University of Science and Technology, Changchun - 130022, China \\
$^2$ College of Physics and electronic information, Baicheng Normal University, Baicheng - 137000, China\\
$^{3}$ Quantum Information Reserch Center, Shangrao Normal University, Shangrao - 334001, China\\}
\email{*yaozh@cust.edu.cn}

\begin{abstract}
The conventional photon blockade for high-frequency mode is investigated in a two-mode second-order nonlinear system embedded with a two-level atom. By solving the master equation and calculating the zero-delay-time second-order correlation function $g^{(2)}(0)$, we obtain that strong photon antibunching can be achieved in this scheme. In particular, we find that by increasing the linear coupling coefficient of the system, a perfect blockade region will be formed near the zero second-order nonlinear coupling coefficient. Similarly, by increasing the nonlinear coupling coefficient of the system, the perfect blockade zone will appear. And this scheme is not sensitive to the reservoir temperature, both of which make the current system easier to implement experimentally.
\end{abstract}






\section{Introduction}
In recent years, the field of cavity optomechanics has received extensive attention. The interaction of light and matter between cavity modes and the influence of mechanical motion have been extensively studied. After the theory of optical mechanical cooling was put forward in the early stage of mechanical resonators, the progress of optical mechanical experiments in recent years makes the mechanical resonators close to the ground state, which opens up a new way for the quantum application of optical mechanical systems~\cite{1,2,4,5,6,7,8,9,10,11,12}. Recently, the inherent nonlinear coupling of optical cavity optical mechanics system has been studied to realize the generation of non-classical light~\cite{13,14}. And nonclassical light generation is crucial in the fields of quantum optics, quantum information, and quantum computation~\cite{15,16,17,18}. The photon blockade (PB) effect is considered to be an important mechanism to generate nonclassical lights.
In theory, there are two physical mechanisms for implementing PB. The observation of antibunching based on the first mechanism requires large nonlinearities to change the energy-level structure of the system, which is called the conventional photon blockade (CPB). The CPB was first observed in an optical cavity coupled to a single trapped atom~\cite{20,21}, and it is realize the CPB mainly by filled the high value nonlinear medium in the cavity to form split energy level. Since it was proposed, this CPB became a research topic and is important for avariety of fundamental studies and practical applications~\cite{22,23,24,25,26,27,28,29,30,31,32,33,34,35,36}. Subsequently, CPB was found in many experiments, including quantum optomechanical systems~\cite{39,40,41}, semiconductor microcavity with second-order nonlinear properties~\cite{43}, and cavity quantum electro dynamics system~\cite{44,46}. Besides the first physical mechanisms, the strong photon antibunching is found by Liew and Savona in a photon moleule system~\cite{47}, and the fundamental principle is destructive quantum interference between distinct driven-dissipative path ways~\cite{48,49,50}, which is called the unconventional photon blockade (UPB). The realizations of the UPB have been extensively studied, such as coupled optomechanical systems~\cite{51,52}, a bimodal optical cavity with a quantum dot~\cite{53,54}, symmetric and antisymmetric modes in weakly nonlinear photonic molecules~\cite{55}, bimodal coupled polaritonic cavities~\cite{56}, the single-mode cavities coupled with three-order nonlinearities~\cite{57,58,59,60}, and coupled single-mode cavities with third-order nonlinearities~\cite{61,62}. Recently, it has been found that destructive quantum interference can be used to improve the intensity of CPB, thus improving the purity of the single photon source~\cite{64}. In addition, the researchers also found that CPB and UPB not only coexist in the same quantum system~\cite{65}, but also have the same coefficient region, which is helpful to further understand the physical mechanism behind CPB and UPB. In general, the single photon produced by CPB has a high average photon number, but its purity is low, and the nonlinear requirement of the system is high, while the $g^{(2)}(0)$ produced by UPB is very low, but the brightness is relatively poor. Valle et al. have found that both advantages of UPB and CPB can be brought together via two pumps~\cite{66}.

In this work, we study a system consisting of two spatially overlapped single-mode semiconductor cavities filled with a two-level atom. The optimal condition of strong anti bunching is obtained by analytical calculation. Numerical calculation shows that the system can realize CPB in high frequency mode, and the numerical solution is in good agreement with the analytical solution. In particular, in the current system, we can increase the linear coupling coefficient of the system to achieve another photon blockade region in addition to two perfect blockade points in the region with small nonlinear coupling coefficient. In the same way, we can also increase the nonlinear coupling coefficient to make the system achieve a blockade region in the region with small linear coupling coefficient. Moreover, the scheme is insensitive to reservoir temperature. Therefore, the system can not only achieve PB, but also the adjustable coefficient, which makes the system easier to realize in the experiment.

The manuscript is organized as follows: In Sec.~{\rm 2}, we introduce the physical model and physical mechanism for the system. In Sec.~{\rm 3}, we illustrate the numerical simulation method. In Sec.~{\rm 4}, we show the comparison of numerical and analytical solutions for the CPB. Conclusions are given in Sec.~{\rm 5}.

\section{Model}
\label{sec:2}
\begin{figure}[h]
\centering
\includegraphics[width=9cm]{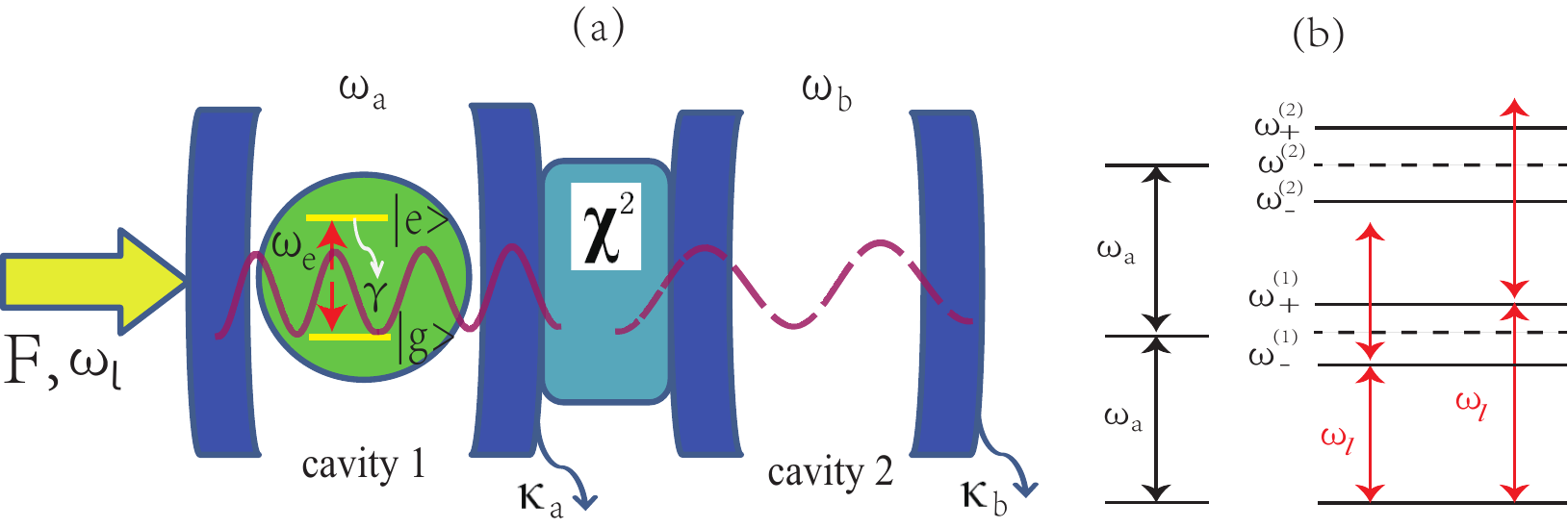}
\caption{ (a) Schematic illustration of a hybrid atom-cavity system,  where cavity 1 which is filled with two-level atom has a high-frequency mode with frequency $\omega_a$ and cavity 2 has a low-frequency mode with frequency $\omega_b$, $\omega_e$ is the frequency of atomic transition from ground state $|g\rangle$ to excited state $|e\rangle$ and $\gamma$ is the atomic spontaneous emission rate, $F$ is the driving strength to the mode $a$ with driving frequency is $\omega_l$, $\kappa$ is the effective decay rate of the two cavities. (b) Schematic level diagram.} \label{fig1}
\end{figure}

The hybrid atom-cavity system is depicted in Fig.~\ref{fig1}(a); this system consists of two single-mode cavities, and the cavitie 1 embedded with a two-level atom (with ground state $|g\rangle$ and excited state $|e\rangle$). The cavitie 2 is a low-frequency cavity with frequencies $\omega_b$, the cavitie 1 is a high-frequency cavity with frequencies $\omega_a$ and the two cavities are coupled via two-order-nonlinear $\chi^{(2)}$ materials that mediates the conversion of the single-photon in cavity 1 into two-photon in cavity 2. External weak drive is the key to realize PB. In this syetem we chose to drive on the cavitie $a$, and the driving frequency is $\omega_l$, driving strength is $F$.

The system Hamiltonian is written as~\cite{pra102,pra97}
\begin{eqnarray}
\hat{H}=\omega_a\hat{a}^{\dag}\hat{a}\
+\omega_e{\sigma}^{\dag}{\sigma}+\omega_b\hat{b}^{\dag}\hat{b}+J(\hat{a}^{\dag}{\sigma}+{\sigma}^{\dag}\hat{a})
+g(\hat{a}^{\dag}\hat{b}^2+\hat{b}^{\dag2}\hat{a})+F(\hat{a}^{\dag}e^{-i\omega_lt}+\hat{a}e^{i\omega_lt}),
\label{01}
\end{eqnarray}
where $\hat{a}(\hat{a}^{\dag})$ and $\hat{b}(\hat{b}^{\dag})$ represents the annihilation (creation) operator of mode $a$ and mode $b$, respectively. The ${\sigma}({\sigma}^{\dag})$ represents the lowering (raising) operator of the two-level atom at the transition frequency $\omega_e$, and $J$ describes the linear coupling strength of the atom and mode $a$, and $g$ expresse the coefficient of second-order nonlinear interactions, which can be derived from the $\chi^{(2)}$ nonlinearity as~\cite{32}
\begin{eqnarray}
g=D\varepsilon_0(\frac{\omega_a}{2\varepsilon_0})\sqrt{\frac{\omega_b}{2\varepsilon_0}}\int d\textbf{r}\frac{\chi^{(2)}(\textbf{r})}{[\varepsilon(\textbf{r})]^{3/2}}\alpha_a^2(\textbf{r})\alpha_b(\textbf{r}).
\label{02}
\end{eqnarray}
And the $\varepsilon_0$ is the vacuum permittivity, $\varepsilon_r$ is the relative permittivity, and $\alpha_a(\textbf{r})$ and $\alpha_b(\textbf{r})$ are the wave functions for mode $a$ and mode $b$, respectively.
In order to study the dynamics of the system, we turn to a rotation framework at the laser frequency $\omega_l$, which leads to a rotated Hamiltonian as
\begin{eqnarray}
\hat{H}_{eff}&=&\Delta_a\hat{a}^{\dag}\hat{a}\
+\Delta_e{\sigma}^{\dag}{\sigma}+\Delta_b\hat{b}^{\dag}\hat{b}+J(\hat{a}^{\dag}{\sigma}+{\sigma}^{\dag}\hat{a})
+g(\hat{a}^{\dag}\hat{b}^2+\hat{b}^{\dag2}\hat{a})+F(\hat{a}^{\dag}+\hat{a}),
\label{01}
\end{eqnarray}
where $\Delta_a=\omega_a-2\omega_l$ and $\Delta_b=\omega_b-\omega_l$ are the cavity detunings, and $\Delta_e=\omega_e-2\omega_l$ is the atom detuning with respect to the driving frequency $\omega_l$. In the closed space formed by the basises $|g,1,0\rangle$, $|e,0,0\rangle$ and $|g,0,2\rangle$, the effective Hamiltonian can be written as a matrix, that is,
\begin{eqnarray}
\tilde{H}=
\begin{bmatrix}
\Delta_a & J & \sqrt{2}g\\
 J & \Delta_e & 0\\
 \sqrt{2}g & 0 & 2\Delta_b
\end{bmatrix}, \label{05}
\end{eqnarray}
The basises we have chosen is in the form of $|g(e),m,n\rangle$, where $g(e)$ means that the atom is in the ground state(excited state), and $m$ and $n$ are photon numbers of modes in cavity 1 and cavity 2, respectively. And the driving terms have been neglected by assuming weak driving. According to the coupling relation of the system and for the convenience of calculation, we make the detunings of the system satisfy the relation of  $\Delta_a=\Delta_e=\Delta$ and $\Delta_b=\frac{1}{2}\Delta_a$ in the following discussion~\cite{pra100}. From diagonalization of the above matrix, and we can get the eigenfrequencies $\xi^{(1)}_{+}$ and $\xi^{(1)}_{-}$ as
\begin{eqnarray}
\xi^{(1)}_{\pm}&=&\Delta\pm\sqrt{2g^{2}+J^{2}}.
\label{06}
\end{eqnarray}
Then we can get the corresponding excitation frequencies $\xi^{(1)}_{\pm} =\sigma^{(1)}_{\pm}+\omega_l$. The schematic level diagram is shown in Fig.~\ref{fig1}(b). We can see that when the driving frequency acting on cavity $a$ is resonant with excitation frequencies, that is, $\omega_{\pm}^{(1)}=\omega_l$, the single-photon probability will increase dramatically. Then we can get the analytical conditions of CPB, which can be written as
\begin{eqnarray}
\sqrt{2}g=\pm\sqrt{\Delta^{2}-J^{2}}.
\label{07}
\end{eqnarray}
In order to realize CPB, the weak driving strength and the nonlinear coupling strength $g\gg\kappa$ must to be satisfied at the same time, the coupling strength $g\gg\kappa$ can ensure a large energy level splitting, and normally we choose to driving on the blocking pattern, which will facilitate the formation of anti-clustering of photons. Satisfy the above conditions, when other parameters are appropriate, the single photon blockade will occur in the mode $a$.

\section{Numerical solution}
\label{sec:3}
The dynamics of this open quantum system is governed by the master equation, i.e.,
\begin{eqnarray}
\frac{\partial\hat{\rho}}{\partial t}&=&-i[\hat{H},\rho]+\frac{\kappa_a}{2}(\bar{n}_{th}+1)(2\hat{a}\hat{\rho}\hat{a}^\dag+\frac{1}{2}\hat{a}^\dag\hat{a}\hat{\rho}+\frac{1}{2}\hat{\rho}\hat{a}^\dag\hat{a})\nonumber\\
&&+\frac{\gamma}{2}(\bar{n}_{th}+1)(2{\sigma}{\rho}{\sigma}^\dag+\frac{1}{2}{\sigma}^\dag{\sigma}\hat{\rho}+\frac{1}{2}{\rho}{\sigma}^\dag{\sigma})\nonumber\\
&&+\frac{\kappa_b}{2}(\bar{n}_{th}+1)(2\hat{b}\hat{\rho}\hat{b}^\dag+\frac{1}{2}\hat{b}^\dag\hat{b}\hat{\rho}+\frac{1}{2}{\rho}\hat{\sigma}^\dag{\sigma})\nonumber\\
&&+\frac{\kappa_a}{2}\bar{n}_{th}(2\hat{a}^\dag\hat{\rho}\hat{a}+\frac{1}{2}\hat{a}\hat{a}^\dag\hat{\rho}+\frac{1}{2}\hat{\rho}\hat{a}\hat{a}^\dag)\nonumber\\
&&+\frac{\gamma}{2}\bar{n}_{th}(2{\sigma}{\rho}{\sigma}^\dag+\frac{1}{2}{\sigma}^\dag{\sigma}\hat{\rho}+\frac{1}{2}\hat{\rho}{\sigma}^\dag{\sigma})\nonumber\\
&&+\frac{\kappa_b}{2}\bar{n}_{th}(2\hat{b}^\dag\hat{\rho}\hat{b}+\frac{1}{2}\hat{b}\hat{b}^\dag\hat{\rho}+\frac{1}{2}\hat{\rho}\hat{b}\hat{b}^\dag) ,\nonumber\\
\label{08}
\end{eqnarray}
where $\kappa_a$ and $\kappa_b$ describe the decay rates of cavities $a$ and $b$, respectively. In order to ensure the generality of the model and the convenience of calculation, we make the relationship between $\kappa_a=\kappa_b=\kappa$ always satisfied in the later calculation. The $\gamma$ describe atomic spontaneous emission rate, at the same time, in the current system, we make $\gamma$ negative and satisfy the relationship of $\gamma=\kappa$~\cite{pra100}. In fact, $\kappa=\kappa_{out}-F$ can be changed by the value of the round trip energy gain $F$, where $\kappa_{out}$ is an intrinsic loss rate. $\kappa > 0$ is the loss and $\kappa < 0$ is the gain, which depends on the relationship between $\kappa_{out}$ and $F$. Actually, the system model of $\kappa < 0$ has been widely studied~\cite{z1,z2}, and has been implemented in recent experiments~\cite{z3,z4}.
$\bar{n}_{th}$ is the mean number of thermal photons, $\kappa_B$ is the Boltzmann constant, and $T$ is the reservoir temperature at thermal equilibrium. The statistic properties of photons will be described by the zero-delay-time second-order correlation function $g^{(2)}(0)$ in steady state, so, we just have to through setting $\partial\hat{\rho}/\partial t=0$ to solve the main equation for the steady-state density operator $\hat{\rho}_s$. In this system, we will analyze the CPB in high-frequency mode $a$, and the statistical properties of photons will be described by zero-delay-time correlation function defined by
\begin{eqnarray}
g^{(2)}(0)=\frac{\langle
\hat{a}^{\dag}\hat{a}^{\dag}\hat{a}~\hat{a}\rangle}
{\langle \hat{a}^{\dag}\hat{a}\rangle^2}=\frac{Tr(\rho_{s}\hat{a}^{\dag}\hat{a}^{\dag}\hat{a}~\hat{a})}{Tr(\rho_{s}\hat{a}^{\dag}\hat{a})}, \label{09}
\end{eqnarray}
The $g^{(2)}(0)<1$ corresponds to the sub-Poissonian statistics. Through this correlation function, we can know whether single photon blocking occurs in high frequency mode $a$. Here we can express the brightness as the average photon number $N=\langle \hat{a}^{\dag}\hat{a} \rangle=[Tr(Tr(\rho_{s}\hat{a}^{\dag}\hat{a})]$, and to achieve a single photon source, we must ensure $g^{(2)}(0) \ll 1$ and sufficient brightness.

\section{Comparison of numerical simulation and analytical conditions}
\label{sec:4}
\begin{figure}[h]
\centering
\includegraphics[width=9cm]{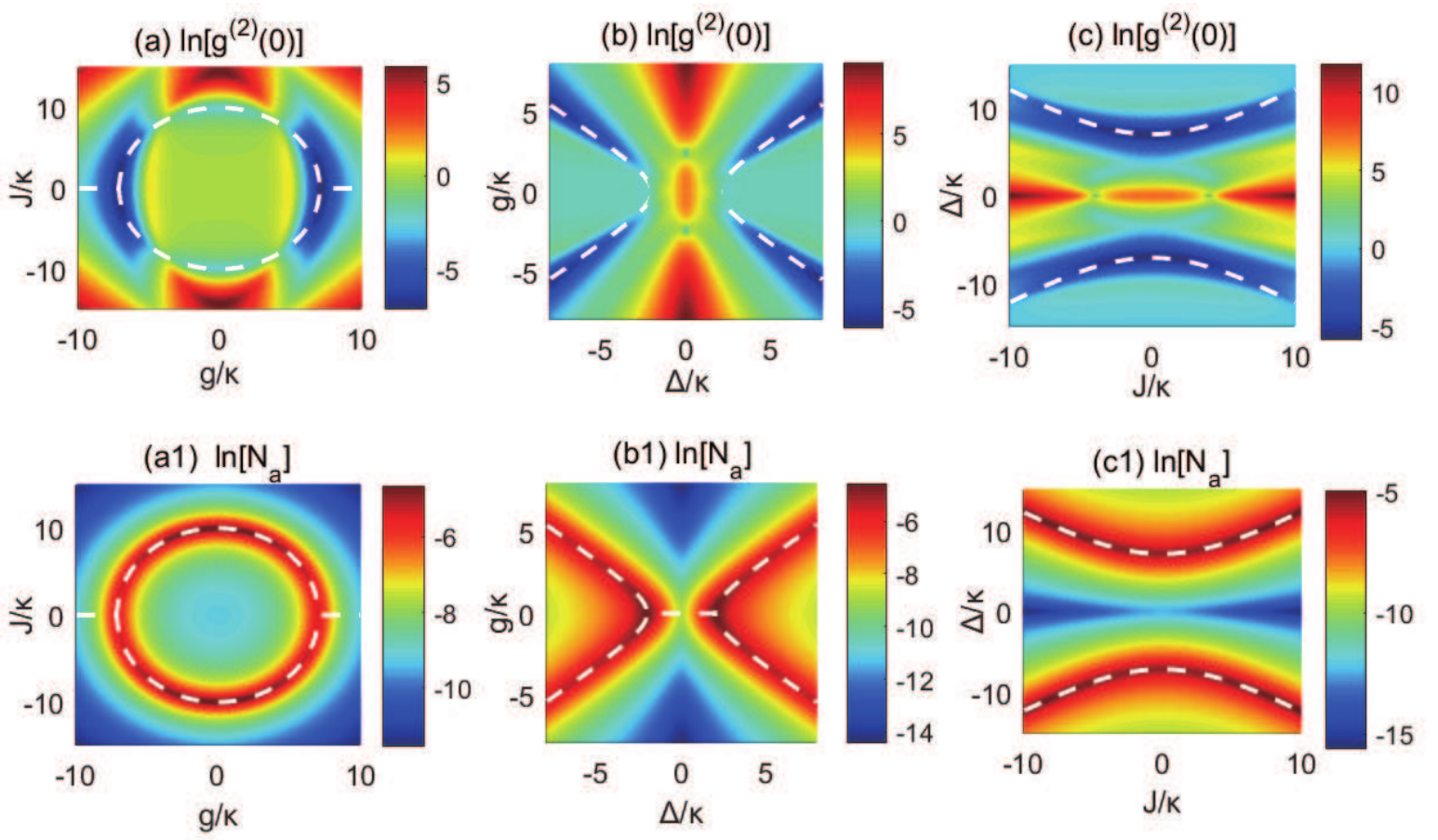}
\caption{(Color online)(a) and (a1)logarithmic plot of the $g^{(2)}(0)$ and average photon number $N_a$ as functions of $g/\kappa$ and $J/\kappa$ for $F_a/\kappa=0.1$, $\bar{n}_{th}=0$ and $\Delta/\kappa=10$. (b) and (b1) logarithmic plot of the $g^{(2)}(0)$ and average photon number $N_a$ as functions of $\Delta/\kappa$ and $g/\kappa$ for $F/\kappa=0.1$, $\bar{n}_{th}=0$ and $J/\kappa=2$. (c) and (c1) logarithmic plot of the $g^{(2)}(0)$ and average photon number $N_a$ as functions of $J/\kappa$ and $\Delta/\kappa$ for $F/\kappa=0.1$, $\bar{n}_{th}=0$ and $g/\kappa=5$. In all the figures (a), (b), (c), (a1), (b1) and (c1) the white dotted line denotes the optimal conditions of CPB shown in Eq.~(\ref{07}). All parameters are in units of $\kappa$ in this paper.}
\label{fig2}
\end{figure}

Now, let us to study the CPB by the numerical simulation, and which compared it with optimal analytic condition show in Eqs.~(\ref{07}).
In order to represent the CPB effect, In Fig.~\ref{fig2} we plot of the $g^{(2)}(0)$ and average photon number $N_a$ versus the system parameters to show the results in a truncated Fock space. And the Hilbert spaces are truncated to two dimensions for the two level atom and the Hilbert spaces are truncated to five dimensions for the modes a and b respectively. In panels (a), (a1), (b), (b1), (c) and (c1) we numerically study the CPB effect under the zero temperature ($\bar{n}_{th}=0$). First of all, we logarithmic plot of the $g^{(2)}(0)$ as a function of $J/\kappa$ and $g/\kappa$ for mode $a$ in panels (a), the other parameters are $F_a/\kappa=0.1$ and $\Delta\kappa=10$. The numerical results show that, the CPB can occur in this system, where the valleys in $g^{(2)}(0)< 1$ corresponds to the strong photon antibunching, the shape of which similar to a circle in the region $g\gg \kappa$. However, the strongest photon anti clustering region in the panels (a) appears in the region where the absolute value of $g$ is larger, which is consistent with the physical mechanism that the system needs larger second-order nonlinearity to achieve the CPB. At the same time, the PB phenomenon also appears in the region where the absolute value of nonlinear coefficient $g$ is small and the value of linear coupling coefficient $J$ is large. The results show that the system can realize PB in the region where the nonlinear coupling strength is small by increasing the linear coupling coefficient of the system. In the following numerical discussion, we will continue to discuss the property that the system can achieve PB in different regions by adjusting its coupling coefficient. The analytic solution is denoted by the white dotted line, which is agree well with numerical simulations. In panels (b), we plot the $g^{(2)}(0)$ as a function of $g/\kappa$ and $\Delta/\kappa$, and we set the $F_a/\kappa=0.1$ and $J/\kappa=2$. The numerical results show that CPB effects can also occur, the shape of which looks like a hyperbola in the region $g\gg \kappa$. At the same time, we find that the blockade effect increases with the detuning $\Delta$, which is consistent with the analytical solution shown in Eq.~(\ref{06}). In panels (c), we plot the $g^{(2)}(0)$ as a function of $\Delta/\kappa$ and $J/\kappa$, and we set the $F_a/\kappa=0.1$ and $g/\kappa=5$. Consistent with the results of panels (b), the photon blockade can be realized in the region with larger detuning $\Delta$, and the numerical solution is consistent with the analytical solution.

In order to further discuss the photon anti clustering effect of the system, in panels (a1) we plot of the average photon number $N_a$ as a function of $J/\kappa$ and $g/\kappa$, and in panels (b1) we plot of the average photon number $N_a$ as a function of $g/\kappa$ and $\Delta/\kappa$, and in panels (c1) we plot of the average photon number $N_a$ as a function of $\Delta/\kappa$ and $J/\kappa$, while the coordinates and parameters are same with panels (a), (b) and (c), respectively. The brightness is defined as average photon number $N_a=\langle \hat{a}^{\dag}\hat{a}\rangle$, which can obtained be by numerically solving the master equation. Compared with panel (a), (b) and (c), the large average photon number region corresponds to the strong photon antibunching region, which indicates that the system can generate more $a$ mode photons in the photon antibunching region, thus increasing the possibility of realizing single photon source. In general, under the weak driving condition, the single-photon probability is far larger than the two-photon probability, i.e., $P_{10}\gg P_{02}$, where $P_{10}$ denotes the probability that there are one photon in mode $a$, $P_{02}$ denotes the probability that there is two photons in mode $b$. However, the single excitation resonance with the Eq.~(\ref{06}) makes the $P_{10}\gg P_{02}$, the $P_{10}$ increasing dramatically due to the single excitation resonance, which makes the CPB occurs in mode $a$.

\begin{figure}[h]
\centering
\includegraphics[width=9cm]{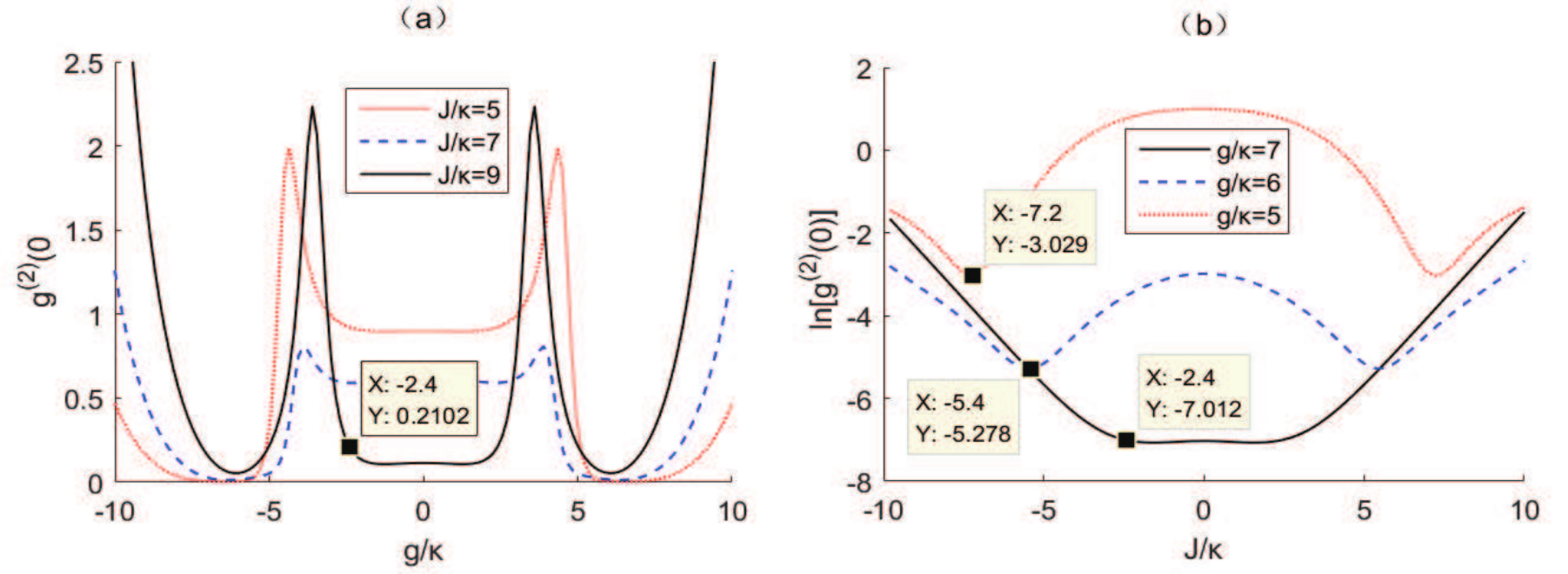}
\caption{(Color online) (a) Under the different linear coupling coefficient $J/\kappa$, we plot the $g^{(2)}(0)$ vs the second-order nonlinear interaction strength $g/\kappa$, with $F/\kappa=0.1$ and $\Delta/\kappa=10$. (b) Under the different nonlinear coupling coefficient $g/\kappa$, logarithmic plot the $g^{(2)}(0)$ vs the $J/\kappa$, with $F/\kappa=0.1$ and $\Delta/\kappa=10$.}
\label{fig3}
\end{figure}

Next, we discuss the influence of the linear and nonlinear coupling coefficients on the realization of PB in the system. In Fig.~\ref{fig3}(a) we plot the $g^{(2)}(0)$ vs the second-order nonlinear interaction strength $g/\kappa$, under the different linear coupling coefficient $J/\kappa$, with $F/\kappa=0.1$ and $\Delta/\kappa=10$. In the black solid line we set $J/\kappa=9$, in the blue dotted line $J/\kappa=7$, in the red point line $J/\kappa=5$. According to the results in the Fig.~\ref{fig3}(a), we find that when $J/\kappa=5$, there are two perfect blockade points at the position satisfying the analytic condition, and when $J/\kappa=7$ is similar to the case of $J/\kappa=5$, only two perfect blockade points appear. With the further increase of $J/\kappa$, when $J/\kappa=9$, not only a perfect blockade point appears at the position where the blockade condition is satisfied, but also a blockade region appears between $g=\pm2.4$. The results show that we can achieve photon blockade in the region with small nonlinear coefficient by increasing the linear coupling coefficient, which increases the possibility of experimental implementation of the system. In order to further discuss the relationship between the two coupling coefficients. In Fig.~\ref{fig3}(b) we logarithmic plot the $g^{(2)}(0)$ vs the $J/\kappa$, under the different nonlinear coupling coefficient $g/\kappa$, with $F/\kappa=0.1$ and $\Delta/\kappa=10$. In the black solid line we set $g/\kappa=7$, in the blue dotted line $g/\kappa=6$, in the red point line $g/\kappa=5$. The results show that under different values of $g/\kappa$, the perfect blockade points is formed at the position satisfying the analytic conditions, and the blockade effect is enhanced with the increase of the nonlinearity of the system. At the same time, when $g/\kappa=7$, a larger perfect blockade zone is formed between $J/\kappa=\pm2.4$. According to the above numerical results, we find that the current system is adjustable in realizing the photon blockade effect. We can achieve a perfect blockade region by increasing the linear or nonlinear coupling coefficient, which greatly increases the experimental feasibility of the system.

\begin{figure}[h]
\centering
\includegraphics[width=9cm]{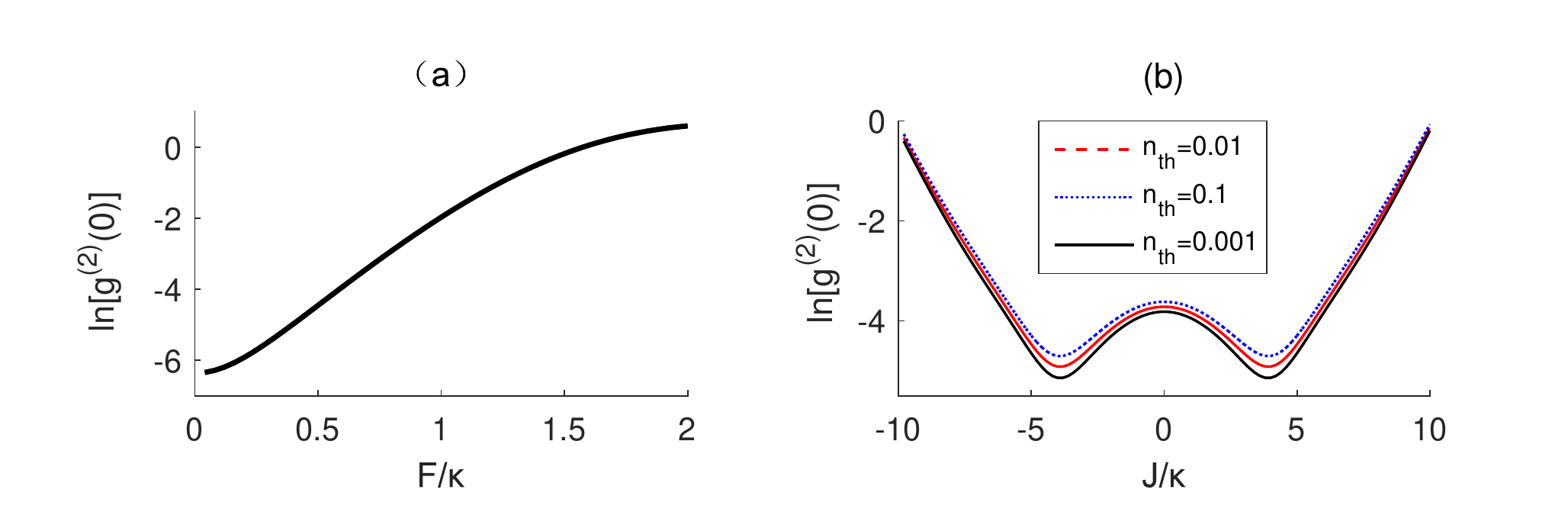}
\caption{(Color online) (a) Logarithmic plot of the $g^{(2)}(0)$ as functions of $F/\kappa$ for $\Delta/\kappa=10$, $\bar{n}_{th}=0$, $J/\kappa=6$ and $g/\kappa=4\sqrt{2}$. (b) Under the different number of thermal photons $\bar{n}_{th}$, we logarithmic plot the $g^{(2)}(0)$ vs the $J/\kappa$, with $F/\kappa=0.1$, $\Delta/\kappa=8$ and $g/\kappa=5$, and in the black solid line we set $\bar{n}_{th}=0.001$, in the red dotted line $\bar{n}_{th}=0.01$ and in the blue point line $\bar{n}_{th}=0.1$.} \label{fig4}
\end{figure}

Next, we discuss the influence of the driving factor on the CPB, in Fig.~\ref{fig4}(a) we logarithmic plot of the $g^{(2)}(0)$ as functions of $F/\kappa$ for $\Delta/\kappa=10$, $\bar{n}_{th}=0$, $J/\kappa=6$ and $g/\kappa=4\sqrt{2}$.  As seen in the Fig.~\ref{fig4}(a), when the value of driving strength $F/\kappa$ increases, the value of $g^{(2)}(0)$ tends to be 1, that is, the CPB phenomenon disappears. Therefore, in order to achieve CPB, the system must meet the condition of weak driving. In the Fig.~\ref{fig4}(b), we plot the $g^{(2)}(0)$ as a function of $J/\kappa$, under the different number of thermal photons $\bar{n}_{th}$, where $F/\kappa=0.1$, $\Delta/\kappa=8$ and $g/\kappa=6$, in the black solid line we set $\bar{n}_{th}=0.001$, in the red dotted line $\bar{n}_{th}=0.01$, and in the blue point line $\bar{n}_{th}=0.1$, the results show that the strongest CPB point appears on $J/\kappa=\pm3.75$, just as predicted. Moreover, when $\bar{n}_{th}$ changes in a large ranges, the PB does not change significantly, which indicate that this scheme is not sensitive to the change of the reservoir temperature, that make the system easier to implement experimentally.

\section{Conclusions}
\label{sec:5}
In summary, we have investigated the CPB for high-frequency mode $a$ in a two-mode second-order nonlinear system. And in this system, a two-level atom is embedded in a high frequency cavity by means of linear coupling, at the same time it is coupled to the low-frequency cavity by second-order nonlinearity. By analytic calculation, we obtain the optimal condition for strong antibunching, and through numerical calculation, we find that the system considered here can realize CPB in the high-frequency mode, and the numerical solution is in good agreement with the analytical solution. Numerical results show that by adjusting the coupling coefficient in the system, the current system can obtain a larger blockade region besides the ideal blockade point. In particular, the nonlinearity of the system can be reduced by increasing the linear coupling of the system, which not only increases the range of optional parameters of the system, but also increases the possibility of realizing the system in experiment. In addition, in order to achieve photon blocking, the system must satisfy the weak driving condition, and the system is insensitive to the change of ambient temperature.
\section*{Acknowledgment}
This work is supported by the National Natural Science Foundation of China with Grants No. 11647054, the Science and Technology Development Program of Jilin province, China with Grant No. 2018-0520165JH, the Jiangxi Education Department Fund under Grant No. GJJ180873.
\section*{Disclosures:} The authors declare no conflicts of interest.\\

\end{document}